\documentclass[doublecol,linenumbers]{epl2} 
\usepackage{graphicx}
\usepackage{subfig}

\title{Otto Engine:  Classical and Quantum Approach }
\shorttitle{Otto Engine:  Classical and Quantum Approach } 

\author{Francisco J. Pe\~na\inst{1},Oscar Negrete\inst{1,2},Natalia Cort\'{e}s\inst{1} and Patricio Vargas\inst{1,2}}
\shortauthor{F. J. Pe\~na \etal}

\institute{                    
  \inst{1} Departamento de F\'isica, Universidad T\'ecnica Federico Santa Mar\'ia, Casilla 110-V 2390123 Valpara\'iso, Chile. \
  
  \inst{2} Centro para el Desarrollo de la Nanociencia y la Nanotecnolog\'ia, 8320000 Santiago, Chile.
}


\abstract{In this paper, we analyze the total work extracted and the efficiency of the magnetic Otto cycle in its classic and quantum versions.  As a general result, we found that the work and efficiency of the classical engine is always greater than or equal to that of its quantum counterpart independent of the working substance. In the classical case, this is due to the fact that the working substance is always in thermodynamic equilibrium at each point of the cycle, maximizing the energy extracted in the adiabatic paths. We apply this analysis to the case of a two-level system, finding that the work and efficiency in Otto's quantum and classical cycle are identical, regardless of the working substance, and we obtain similar results for a multilevel system where a linear relationship between the spectrum of energies of the working substance and the external magnetic field is fulfilled. Finally, we show an example of a three-level system in which we compare two zones in the entropy, temperature and magnetic field diagram to find which is the most efficient when performing a thermodynamic cycle. This work provides a practical way to look for temperature and magnetic field zones in the entropy diagram that can maximize the power extracted from Otto's magnetic engine.}

\begin{document}

\maketitle

\section{Introduction}

The classical standard non-magnetic Otto cycle is widely used in present-day technology, as it is the thermodynamic cycle most commonly found in automobiles engines. This cycle consists of two classical isochoric and two classical adiabatic processes. In the isochoric processes, the system interacts with either of two thermal reservoirs at temperatures $T_{l}$ and $T_{h}$, with $T_{h} > T_{l}$, and each one of these processes is followed by a classical adiabatic process which enables the work extraction.  In this cycle, the results of the efficiency depend on the nature of the working substance (through its energy spectrum), and the contributions of work and heat are separate in its stages, which facilitates the theoretical modelling. These characteristics favor its extension to its quantum version, which has been studied extensively in recent years\cite{Su2016,Kosloff2014,Lutz2017,Lutz2_2017,Biswas2017,Reid2018,Beretta2012,Lutz3_2016,Solfanelli2020,Mehta2017,Leggio2016,Alvarado2017,Hewgill2017,Myers2020,Deffner2020,Kosloff2017}.

Otto's quantum cycle similarly consists of four processes: two quantum isochoric processes and two quantum adiabatic processes. The quantum isochoric process is very similar to the classical one, in the sense that both admits changes in the temperature of the systems trough heat exchange with zero work performed \cite{Quan2009}. In contrast, quantum adiabatic processes are conceptually different from its classical counterpart. The adiabatic process in the classical case can be achieved by rapid expansion and compression over the working substance, to guarantee no heat exchange with the thermal bath. On the contrary, the quantum scenario requires that the energy levels populations of the working substance remain constant for each of the quantum states as the volume or external magnetic field varies, ensuring that entropy remains unchanged.  This implies that the process should be realized in a quasi-static way and, therefore, it must be slow enough to avoid transitions between levels that can be generated while it is carried out.  Deffner and Campbell define this difference \cite{DeffnerCampbell}: ``\textit{Quantum adiabatic process form only a subset of classical adiabatic process}". Therefore, the developing a classical and quantum Otto cycle differs not only in the time of their processes but also in the physical concepts involved. Besides, it is essential to mention that the word ``classical" should not be directly interpreted as the working substance is ``classical." The notion of ``classical" refers here to the four stages developed in the classical Otto cycle are in thermodynamic equilibrium, and consequently the resultant cycle is a reversible process.  Otherwise, the working substance of quantum cycle reaches thermal equilibrium only in two stages of the cycle, setting up an irreversible process.

Given the differences of the classical and quantum process of the adiabatic Otto's stage, two questions naturally arise: How are work performance and efficiency when these two approaches, classical and quantum, are applied to the same working substance?, and what conditions should fulfill the energy spectrum of the working substance in both approaches so that this difference is as small as possible? 

In this work we give a possible solution to these questions using as example an Otto magnetic cycle under the classical and quantum thermodynamics formulation. 
Our first discussion involves a comparison between both thermodynamical approaches to explore why the classical work is greater than its quantum counterpart. We argue that the answer can be regarded as a consequence of a free energy principle for thermal equilibrium systems. As an example we analyze a particular case of a two-level system working under an Otto cycle and we show that the total extracted its equal under both formulations regardless of the levels magnetic dependence. Keep in mind that this result was found imposing a strong probability conservation for the quantum case. Interestingly, this same result can be obtained if on a classical isentropic trajectory the temperature and
the control parameter (volume or external magnetic field) of the cycle are linearly related to each other, giving as consequence a constant population along the process satisfying the conservations requirements of the quantum adiabatic process. Finally, we explore the consequences of the inclusion of more energy levels and how its affects the behavior of thermal quantities with respect the simpler case of just two levels.

\section{Quantum and Classical Magnetic Otto Cycle: Classical and Quantum Work}

To describe the Otto cycle it their classical and quantum formulation, we show a schematic representation of an entropy ($S$) - external field diagram. In our notation, $T$ will refer to temperature and $B$ to external magnetic field (both parameters measured in arbitrary units). The adiabatic and isomagnetic processes are represented in the figure by horizontal lines and vertical ones.

\begin{figure}[!ht]  
\includegraphics[width=0.48 \textwidth]{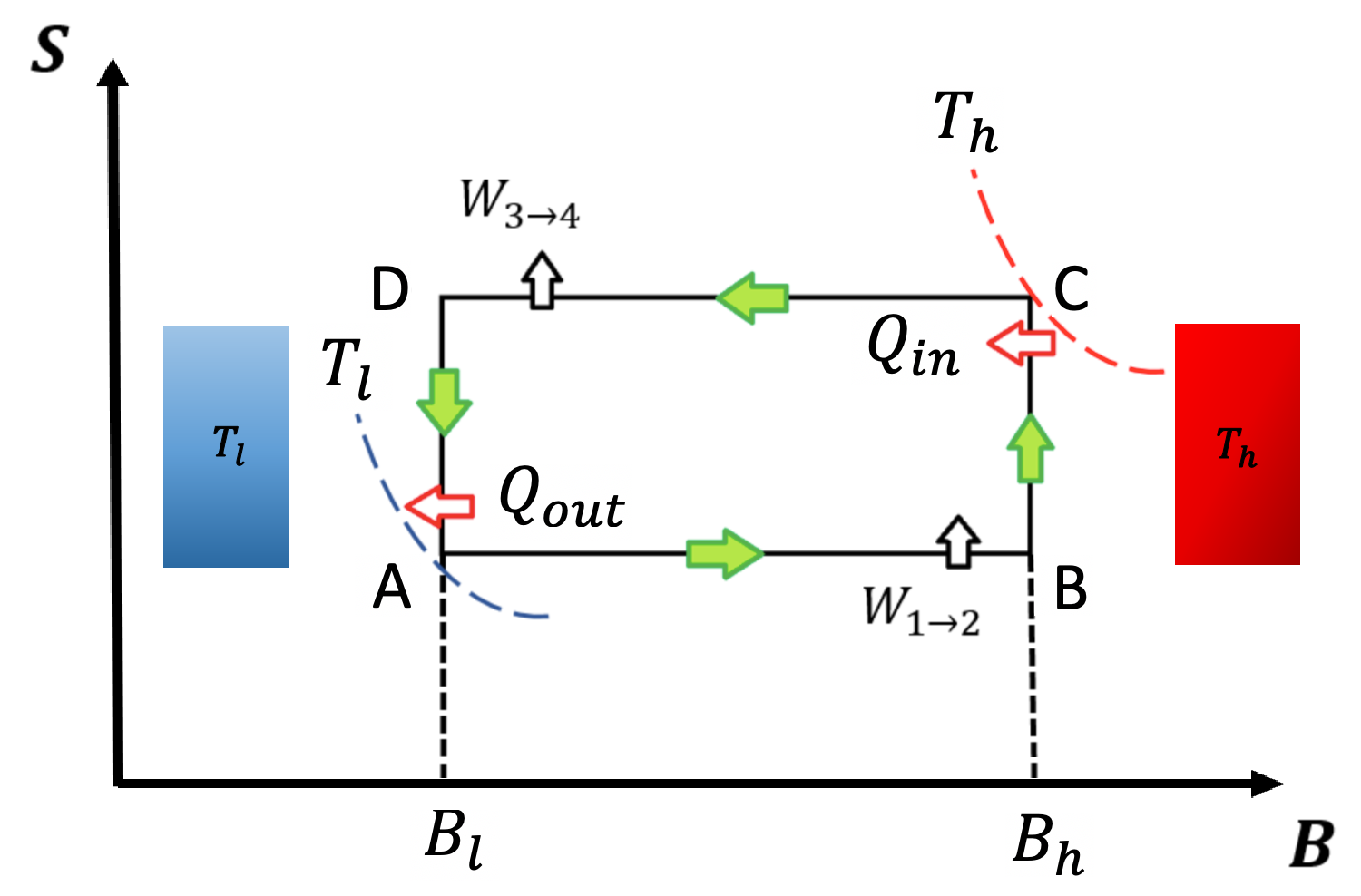}
 \caption{Entropy versus external field diagram for classical and quantum Otto Cycle. The system contacts the thermal reservoirs only in the isomagnetic strokes. At the points $\mathrm{C}$ and $\mathrm{A}$, the working substance reaches the temperatures $T_{h}$ and $T_{l}$, respectively indicated with the isotherms touching the points. For the quantum version, the entropy values $S_{\mathrm{B}}$ and $S_{\mathrm{D}}$  are calculated using the same thermal probabilities as in  points  $ \mathrm{A}$  and $ \mathrm{C}$  to ensure the quantum adiabatic strokes $\mathrm{A} \rightarrow\mathrm{B}$ and $\mathrm{C} \rightarrow\mathrm{D}$.}
 \label{OttoCycle}
 
\end{figure}

The cycle is thermodynamically defined by the two heat reservoirs (hot and cold) and two extreme values of the external field, $B_{h}$ and $B_{l}$. It is usually used that the low temperature stage $(T_{l})$ corresponds to  low external field value $(B_{l})$ and for the high temperature stage $(T_{h})$,  the high external field value ($B_{h}$). This is not true for all working substances \cite{Pena2019,Pena2020} because we need to observe the behaviour of entropy as a function of different temperatures and external field to know which especific values corresponds to each point in the cycle. As example, the entropy behavior in quantum dots \cite{Pena2019,Pena2020} is such that the points where the temperature reservoirs are located are crossed with the fields values. Nevertheless, the conclusions obtained in this work can be applied to this kind of systems regardless of this.

The cycle presented in fig. \ref{OttoCycle} operates following the sequence $\mathrm{A} \rightarrow\mathrm{B}\rightarrow\mathrm{C}\rightarrow \mathrm{D} \rightarrow \mathrm{A}$. The output power is defined as the work done per cycle divided by $\tau$, being $\tau$ the duration of each cycle iteration. It is important to point out that, under this formulation, thermal equilibrium is reached at the thermal reservoirs for both classical and quantum Otto cycle versions. For the first isomagnetic stroke, from  $\mathrm{B} \rightarrow \mathrm{C}$ (see fig. \ref{OttoCycle}), with a hot reservoir at temperature $T_{h}$, where at point $\mathrm{C}$, the working substance reaches the temperature of the hot reservoir. For the second isomagnetic process, from $\mathrm{D}$ to $\mathrm{A}$ (see fig. \ref{OttoCycle}), the system is put in to contact with a cold reservoir at temperature $T_{l}$ up to the working substance reach the same temperature of the cold reservoir. Contrary to the cases of adiabats (classical and quantum case) where the systems are disconnected from the reservoirs, and the external field is varied from $B_{l}$ to $B_{h}$ (process $\mathrm{A}\rightarrow \mathrm{B}$) and vice-versa (process $\mathrm{C}\rightarrow \mathrm{D}$). We use the superscript $q$ for all quantum thermodynamics variables while we use $cs$ for the classical ones. 

In the quantum scenario, the heat absorbed $(Q_{in}^{q})$ and released $(Q_{out}^{q})$ is given by:\cite{Alvarado2017, Quan2009, Pena2019, Pena2020}

\begin{eqnarray}
\label{qin}
Q_{in}^{q}=\sum_{s}E_{s}(B_{h})\left[P^{\mathrm{C}}_{s}(T_{h},B_{h})-P^{\mathrm{B}}_{s}\right],
\end{eqnarray}
\begin{eqnarray}
\label{qout}
Q_{out}^{q}=\sum_{s}E_{s}(B_{l})\left[P^{\mathrm{A}}_{s}(T_{l},B_{l})-P^{\mathrm{D}}_{s}\right],
\end{eqnarray}

where $E_{s}(B_{l(h)})$ is the energy spectrum of the working substance evaluated in the low (high) temperature value in the cycle, $P_{s}$ corresponds to the occupation probabilities along the cycle and the index $s$ represents the different  numbers that characterise a quantum state of the working substance.  To satisfy the adiabatic nature of the process under the quantum formulation, one way is that the occupancy probabilities must satisfy the following conservation condition,

\begin{eqnarray}
\label{adibaticcond}
P^{\mathrm{B}}_{s}=P^{\mathrm{A}}_{s}(T_{l},B_{l}), \quad P^{\mathrm{D}}_{s}=P^{\mathrm{C}}_{s}(T_{h},B_{h}).
\end{eqnarray}

Using these relations, we can define the total work per cycle as:
\begin{eqnarray}
\label{quantumwork}
\mathcal{W}^{q}&=&Q_{in}^{q}+Q_{out}^{q}
\nonumber
\\ &=&\sum_{s}\left(E_{s}(B_{h})-E_{s}(B_{l})\right) \\ 
\nonumber
&& \times  \left[P^{\mathrm{C}}_{s}(T_{h},B_{h})-P^{\mathrm{A}}_{s}(T_{l},B_{l})\right].
\end{eqnarray}

Here we can see one of the main difference between the classical and quantum approach. While in classical thermodynamics we require the system to be in thermal equilibrium at every moment, here we can see that even if the points B and D doesn’t fulfill this condition, the entropy conservation along the adiabatic paths can be made just with only two equilibrium points (A and C). Here we suppose a situation in which the quantum states are orthogonal i.e we have a diagonal density matrix.  With this Ansatz, the entropy, $S$, of the system is simply $S= - \sum_s P_s \ln (P_s)$, which is reduced to thermodynamic entropy when the probabilities are calculated in equilibrium, i.e., where the temperature is defined at each point of the cycle. 

On the other hand, the classical case does not require the strong conservation restriction of the population (\ref{adibaticcond}) giving the possibility of having variations in thermal occupations along the adiabatic pathway. Furthermore, as the systems follow the classical thermodynamic formulation, they are kept in equilibrium in all positions of the $S-B$ diagram and the temperature can be defined at points $B$ and $D$. A possible way to do this is to solve the equation of the total entropy differential $dS(T,B)=0$ to obtain information about the relationship between $T$ and $B$ along the isentropic processes. The first order differential equation is given by:

\begin{eqnarray}
\label{differential}
\frac{dB}{dT}=-\frac{\left(\frac{\partial S}{\partial T}\right)_{B}}{\left(\frac{\partial S}{\partial B}\right)_{T}}.
\end{eqnarray}

Other possibility is simply impose the classical adiabatic condition between the points  A-B an C-D in the form of 

\begin{eqnarray}
\label{equality}
S(T_{l},B_{l})=S(T_{\mathrm{B}},B_{h}), \quad S(T_{h},B_{h})=S(B_{l},T_{\mathrm{D}}).
\end{eqnarray}

Because in the classical case the working substance is always in thermal equilibrium, the internal energy $U(T,B)$ derived from the canonical partition function $(\mathcal{Z}(T,B))$, it is always well defined, i.e. $U = T^2\frac{ \partial \ln(\mathcal{Z})}{ \partial T}$. Accordingly, the incoming heat and released heat can be rewritten for the classical case as 

\begin{eqnarray}
\label{qincla}
Q_{in}^{cs}=U_{\mathrm{C}}(T_{h},B_{h})-U_{\mathrm{B}}(T_{\mathrm{B}},B_{h}),
\end{eqnarray}
\begin{eqnarray}
\label{qoutcla}
Q_{out}^{cs}=U_{\mathrm{A}}(T_{l},B_{l})-U_{\mathrm{D}}(T_{\mathrm{D}},B_{l}).
\end{eqnarray}

According to these two expressions,  the total work extracted in the classical formulation is given by:
\begin{eqnarray}
\label{classicalwork}
\mathcal{W}^{cs}&=& \left[U_{\mathrm{C}}(T_{h},B_{h})+U_{\mathrm{A}}(T_{l},B_{l})\right] \\ 
\nonumber
&-&\left[U_{\mathrm{B}}(T_{\mathrm{B}},B_{h})+U_{\mathrm{D}}(T_{\mathrm{D}},B_{l})\right], 
\end{eqnarray}
where we have separated the points where the working substance comes into contact with thermal reservoirs (points A and C) and the other points assumed in thermal equilibrium for the classical formulation (points B and D). 

On the other hand, we can rewrite the equation (\ref{quantumwork}) associated to the quantum work as follow

\begin{eqnarray}
\mathcal{W}^{q}=\left[U_{\mathrm{C}}(T_{h},B_{h})+U_{\mathrm{A}}(T_{l},B_{l})\right]-\left[U^{*}_{\mathrm{B}} + U^{*}_{\mathrm{D}}\right], 
\label{quantumwork1}
\end{eqnarray}
where $U^{*}_{\mathrm{B}}$ and $U^{*}_{\mathrm{D}}$ are two expected values of energy in non-thermal equilibrium. This is where we can get a first relevant discussion. According to thermodynamics, a system in equilibrium has the minimum value of energy for a given entropy. If this were not so, we can think that we could withdraw energy from the system (for example, in the form of work), keeping the value of the entropy constant, and then we could return this energy to the system in the form of heat \cite{Callen}. This would leave the system with its initial state of energy but would cause the entropy to increase, violating the condition that for a state of equilibrium, the value of the entropy corresponds to a maximum \cite{Callen}. Therefore, inspecting the equations (\ref{classicalwork}) and (\ref{quantumwork1}), we can argue that the quantity $U^{*}_{\mathrm{B}} + U^{*}_{\mathrm{D}}$ is always greater than $U_{\mathrm{B}}(T_{\mathrm{B}},B_{h})+U_{\mathrm{D}}(T_{\mathrm{D}},B_{l})$. Accordingly, we can conclude that classical work will always be greater or equal than quantum work because the two first terms in equations (\ref{classicalwork}) and (\ref{quantumwork1}) are equal. Consequently we can write the condition for the total work extraction 

\begin{eqnarray}
\label{conditionwork}
\mathcal{W}^{cs}\geq \mathcal{W}^{q}.
\end{eqnarray} 

This is a clear example of the robustness of thermodynamics. The result presented in the equation (\ref{conditionwork}) is nothing more than the condition that a reversible total work is always greater than or equal to the irreversible one, i.e.

\begin{eqnarray}
\mathcal{W}_{reversible}\geq \mathcal{W}_{irreversible}.
\end{eqnarray}

In the next subsection, we will show the analysis of two cases. The first corresponds to a two-level system, and the second a three-level system that simulates a graphene quantum dot. In the first case, we find that the quantum and classical formulation extract the same amount of work, and in the second case, the extraction of quantum work is always less than or equal to its classical counterpart.

\section{The case of a two-level system}

In this section, we address the case of a two-level system to guide the discussion towards the classical and quantum comparison of work and efficiency in the Otto magnetic cycle. For this purpose, let us consider a working substance described by two levels of energy which are continuous and differentiable as a function of magnetic field. These energies which we will call $E_{1}(B)$ and $E_{2}(B)$.   As usual, we have the thermal populations: 

\begin{eqnarray}
\label{popul}
P_{1}(T,B)=\frac{e^{-\beta E_{1}(B)}}{\mathcal{Z}(T,B)}, \quad P_{2}(T,B)=\frac{e^{-\beta E_{2}(B)}}{\mathcal{Z}(T,B)},
\end{eqnarray} 

where $\mathcal{Z}(T,B)=e^{-\beta E_{1}}+e^{-\beta E_{2}}$ and $\beta=\frac{1}{T}$. The thermal populations satisfies the normalization condition given by $P_{1}(T,B)+P_{2}(T,B)=1$. On the other hand, the entropy of von Neumann will be defined as ( with $k_{B}=1$)
\begin{equation}
\label{evonneumann}
S=-P_{1}\ln\left(P_{1}\right)-P_{2}\ln\left(P_{2}\right).
\end{equation}

If we develop the derivatives of the entropy in thermal equilibrium ($S(T,B)$) we obtain  the following expression for the behaviour of the field and temperatures along isentropic strokes given by 

\begin{eqnarray}
\label{twolevelgeneral}
\frac{dT}{dB}=\frac{T\left(\frac{d E_{1}(B)}{dB}-\frac{d E_{2}(B)}{dB}\right)}{E_{1}(B)-E_{2}(B)},
\end{eqnarray}
whose trivial solution is given by
\begin{eqnarray}
\label{generalsolution}
T(B)=\mathcal{C}\left(E_{1}(B)-E_{2}(B)\right),
\end{eqnarray}
where $\mathcal{C}$ it is an integration constant. Equation (\ref{generalsolution}) is a general solution for temperature, independent of the behavior of the energy levels upon the external field $B$. On the other hand,  replacing this solution for temperature in the thermal population defined in equation (\ref{popul}), we obtain that the thermal populations become only dependent on the integration constant $\mathcal{C}$ in the form

\begin{eqnarray}
\label{indepopu}
P_{1}(T(B),B)=\frac{1}{1+e^{\mathcal{C}}}, \quad P_{2}(T(B),B))=\frac{e^{\mathcal{C}}}{1+e^{\mathcal{C}}}, 
\end{eqnarray}

\begin{figure}[!ht]
       \includegraphics[width=0.48\textwidth]{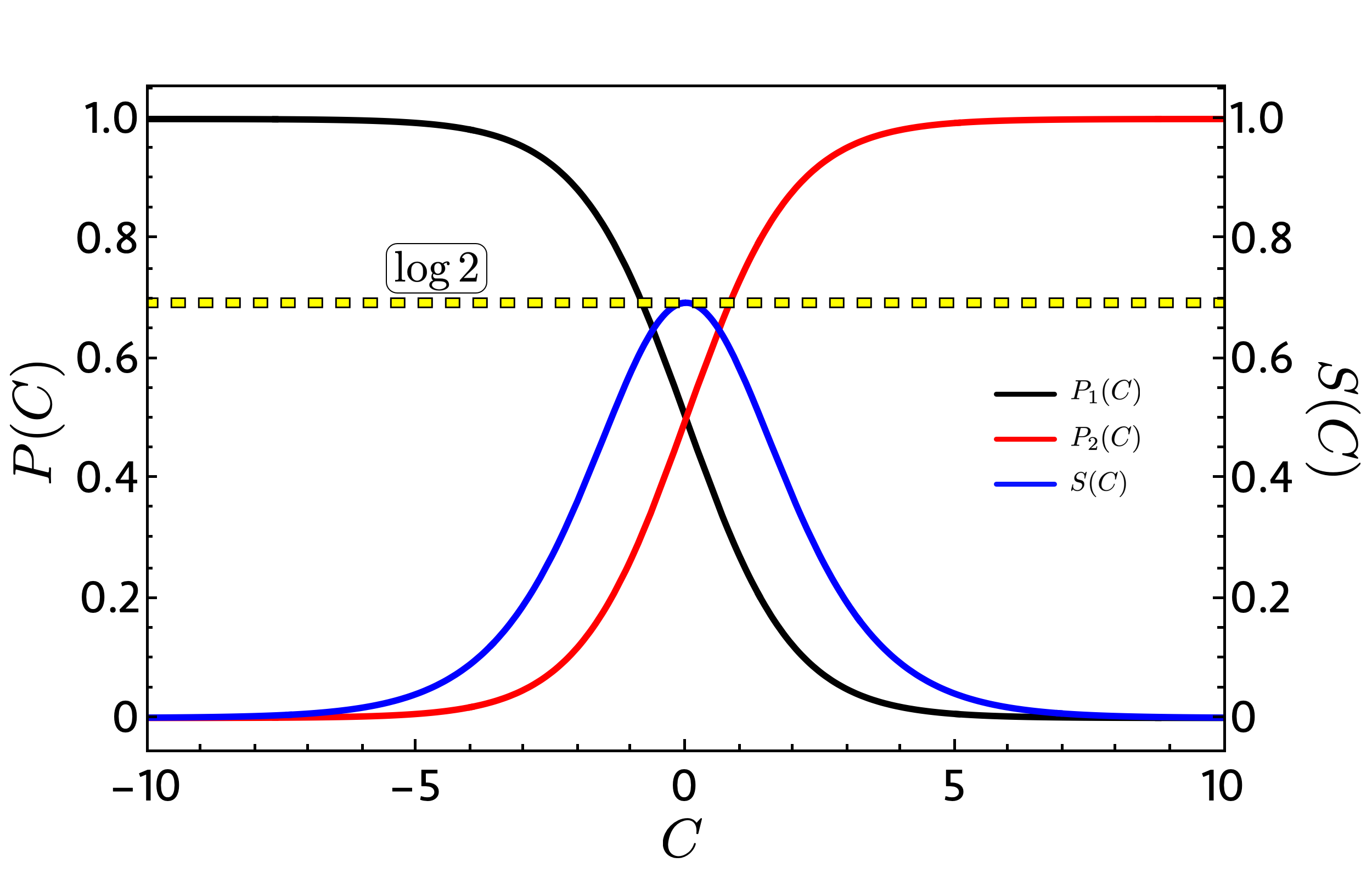}
     \caption{Probabilities $(P)$ and entropy $(S)$ as a function of the integration constant $\mathcal{C}$. We observe that when the integration constant is zero, we obtain the maximum value of the entropy given by $\ln(2)$, because the probabilities take the value of $1/2$ at each level.}
     \label{fig2levelscon}
   \end{figure}

and are presented as a function of the integration constant together with entropy in the fig. (\ref{fig2levelscon}). Consequently, the entropy given by equation (\ref{evonneumann}), is only a function of $\mathcal{C}$, therefore, for a two-level system in the classical approach, we found that the thermal populations in an isentropic stroke become constant. This brings us an immediate consequence: quantum work and classical work become the same because quantum adiabaticity requires to keep the populations constant throughout the process, a situation that is always the case for a two-level system even in its classical treatment as we have shown in this example. Therefore there will be no difference in the mathematical analysis between one and the other approach (only conceptual).

In this same line of discussion, we found another case for $\mathcal{W}$$^{cs}$=$\mathcal{W}$$^{q}$, corresponding to the instance when the working substance in the classical adiabatic strokes satisfies the differential equation in the form

\begin{eqnarray}
\label{propto}
\frac{dT}{dB}\propto\frac{T}{B},
\end{eqnarray}
whose trivial solution corresponds to a linear relation between the magnetic field and temperature, that is $T(B)=\mathcal{C}_{1} B$, where $\mathcal{C}_{1}$ is an integration constant. Take as example a system whose energy levels are mathematically described as $E(B)=(-1)^{j}j B$, where $j$ can take integer values from zero onwards. Thermal populations and the partition function for this system will be defined as:

\begin{eqnarray}
\label{probaparti}
P_{j}=\frac{e^{\frac{(-1)^{j+1} j B}{T}}}{\mathcal{Z}(T,B)}, \quad \mathcal{Z}(T,B)=\sum_{j}e^{\frac{(-1)^{j+1} j B}{T}}.
\end{eqnarray}

It is easy to show that this kind of system in an isentropic stroke has a solution between the variables like in equation (\ref{propto})  (i. e. $T\propto B$).  If we replace this solution in equation (\ref{probaparti}), it is clear that we obtain a constant value for the partition function and, consequently, for the populations of every state. Accordingly, the quantum work and classical work becomes equal in magnitude for the same discussion made before.

\section{The case of a three-level system}

\begin{figure}[!ht]
     \subfloat[ \label{spectrum1}]{%
       \includegraphics[width=0.45\textwidth]{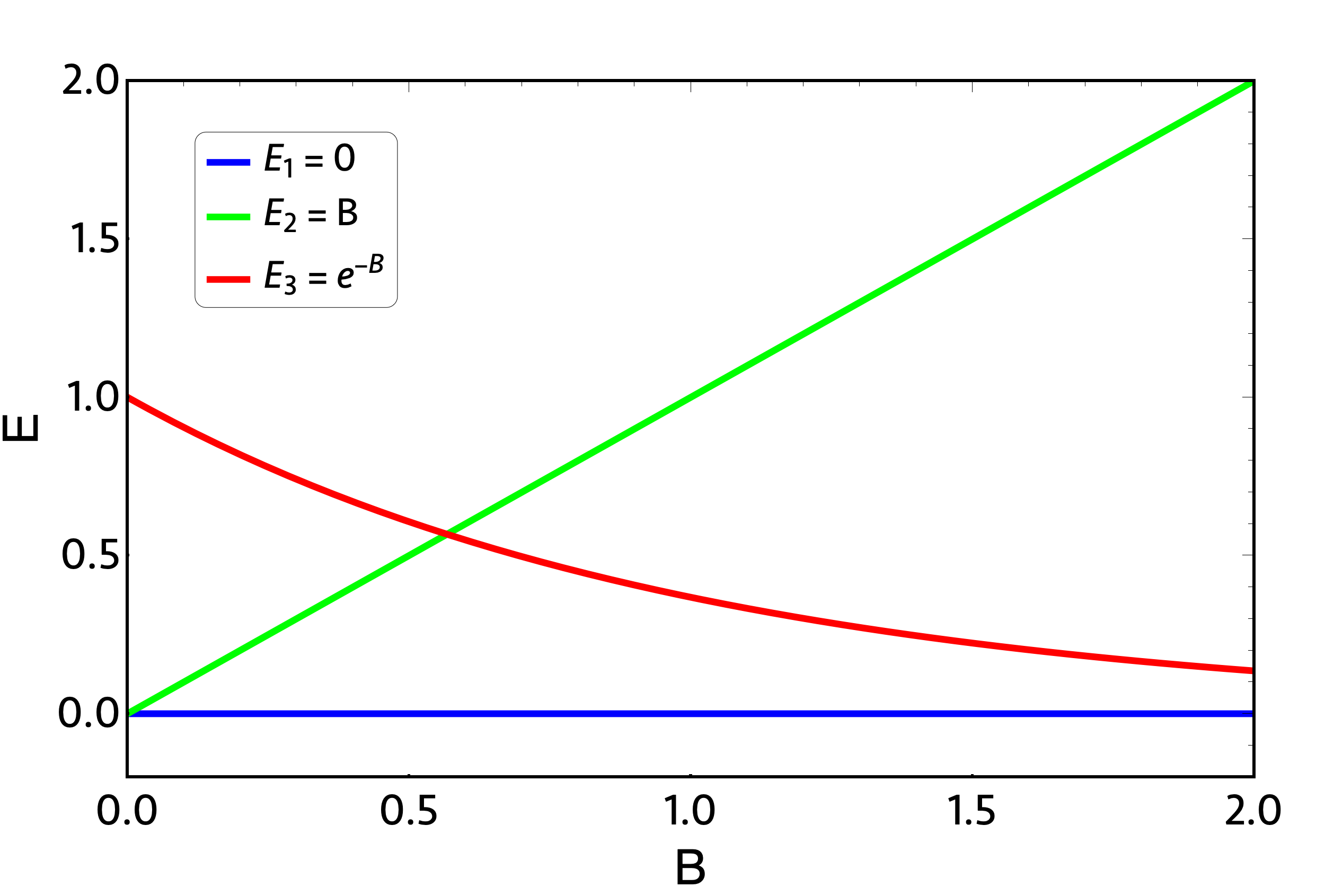}
     }
     \quad
     \subfloat[\label{entropy3-levels}]{%
       \includegraphics[width=0.45\textwidth]{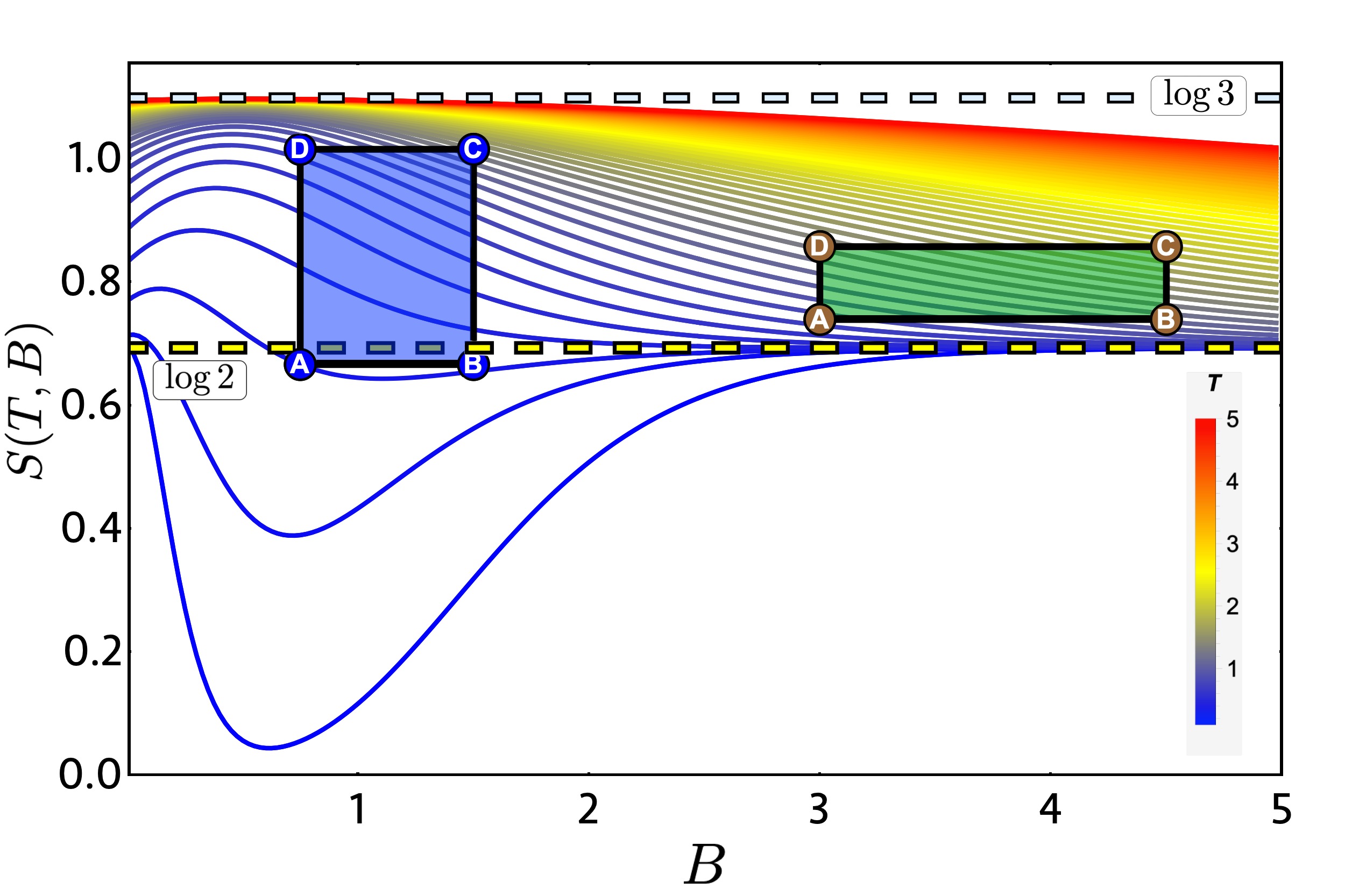}
     }
     \caption{(a) Proposed three-level energy spectrum as a function of external dimensionless magnetic field parameter $B$. (b) Entropy ($S$) versus external field diagram ($B$) in arbitrary units, and the two zones where the Otto cycle in its classical and quantum version is applied. The green rectangle represents the zone 1 and the blue rectangle the zone 2 in the discussions throughout the text.}
     \label{entropy3levels}
   \end{figure}
   
In order to show a case in which the quantum work and the classical work do not coincide, let us take as example the case of a three-level system, in which the energy spectrum is presented in the fig. \ref{entropy3levels}(a) , in arbitrary units. This spectrum of energies simulates (roughly) the case of graphene quantum dot under external magnetic field \cite{Pena2020}. The levels of energy displayed in the fig. (\ref{spectrum1}) are given by $E_{1}=0$, $E_{2}=B$ and $E_{3}=e^{-B}$. This spectrum of energy exhibits a crossing between the levels $E_{2}$ and $E_{3}$ for $B>0$, and is doubly degenerated for zero external magnetic field.

\begin{figure}[!ht]  
\includegraphics[width=0.45\textwidth]{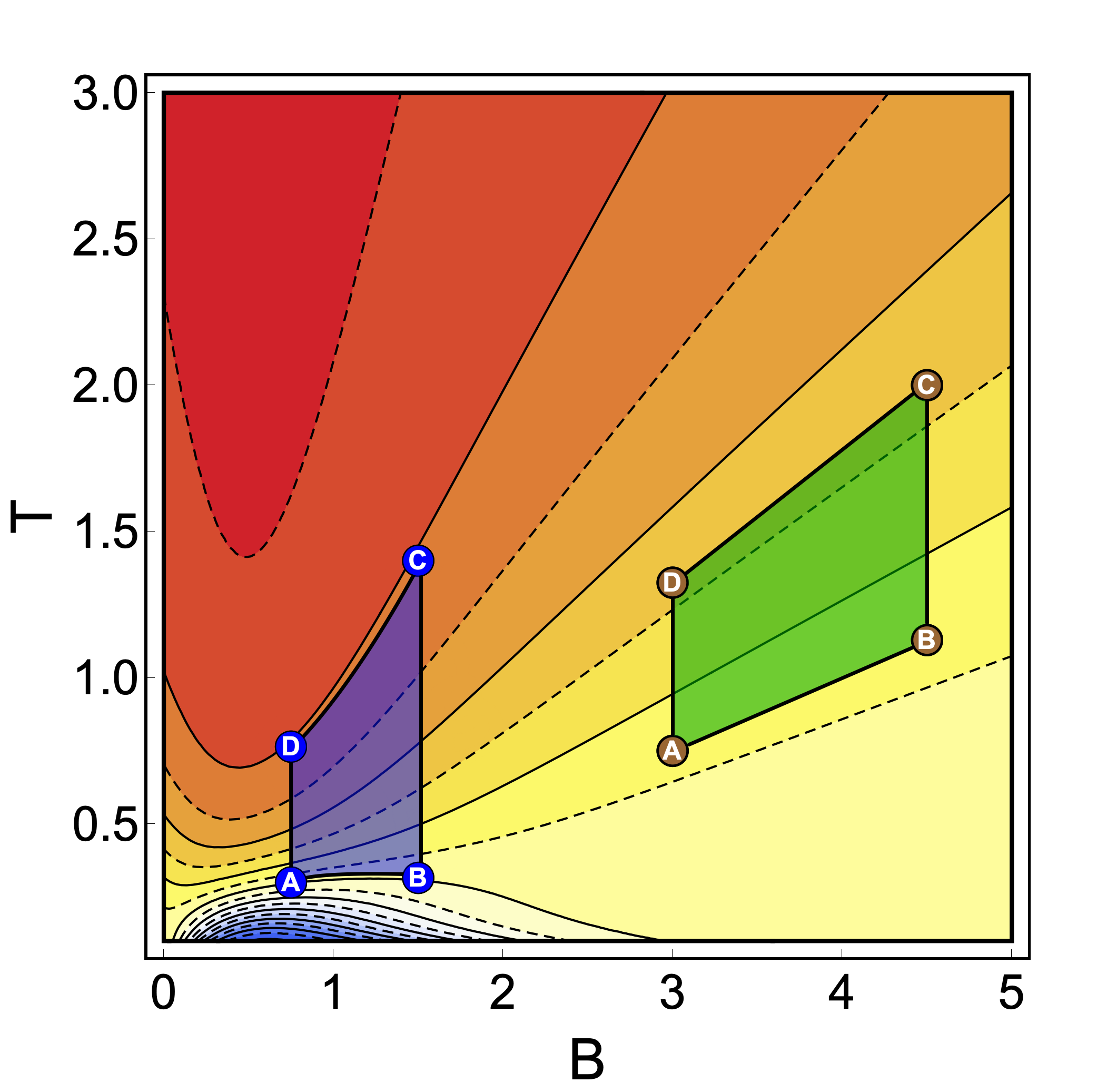}
\caption{ Temperature versus external field diagram obtained when the classical conditions over the isentropic trajectories are applied. From this diagram, the green rectangle (zone 1)  represent a zone where the temperature is linear with the magnetic field, and the blue rectangle (zone 2)  a zone where the temperature in terms of the magnetic field is not linear.}
 \label{quasistatictrajectories}
\end{figure}

The thermodynamics quantities are calculated from the partition function $\mathcal{Z}=1+e^{-\frac{B}{T}}+e^{-\frac{e^{-B}}{T}}$, and the entropy $S$ as a function of external magnetic field is presented in the fig. \ref{entropy3levels}(b). At high temperature and low-external magnetic field region, the entropy tends to $\ln(3)$, this because at high temperature all three energy levels have the same occupation, therefore the same probability. From the same figure, we note some combinations of parameters at low -temperature produces first a decreasing form for the entropy as a function of the external magnetic field and then an increase in its value converging to $\ln(2)$ for high magnetic fields. This is due to the shape of the $E_{3}$ level since for high magnetic fields; it begins to be easier to populate than the $E_{2}$ state that corresponds to a state that grows linear with the value of the external field. Consequently, in a region of a high magnetic field and low temperature, we have only two populated states obtaining the value of $\ln(2)$ for the entropy again.

\begin{figure}[!ht]
       \includegraphics[width=0.48\textwidth]{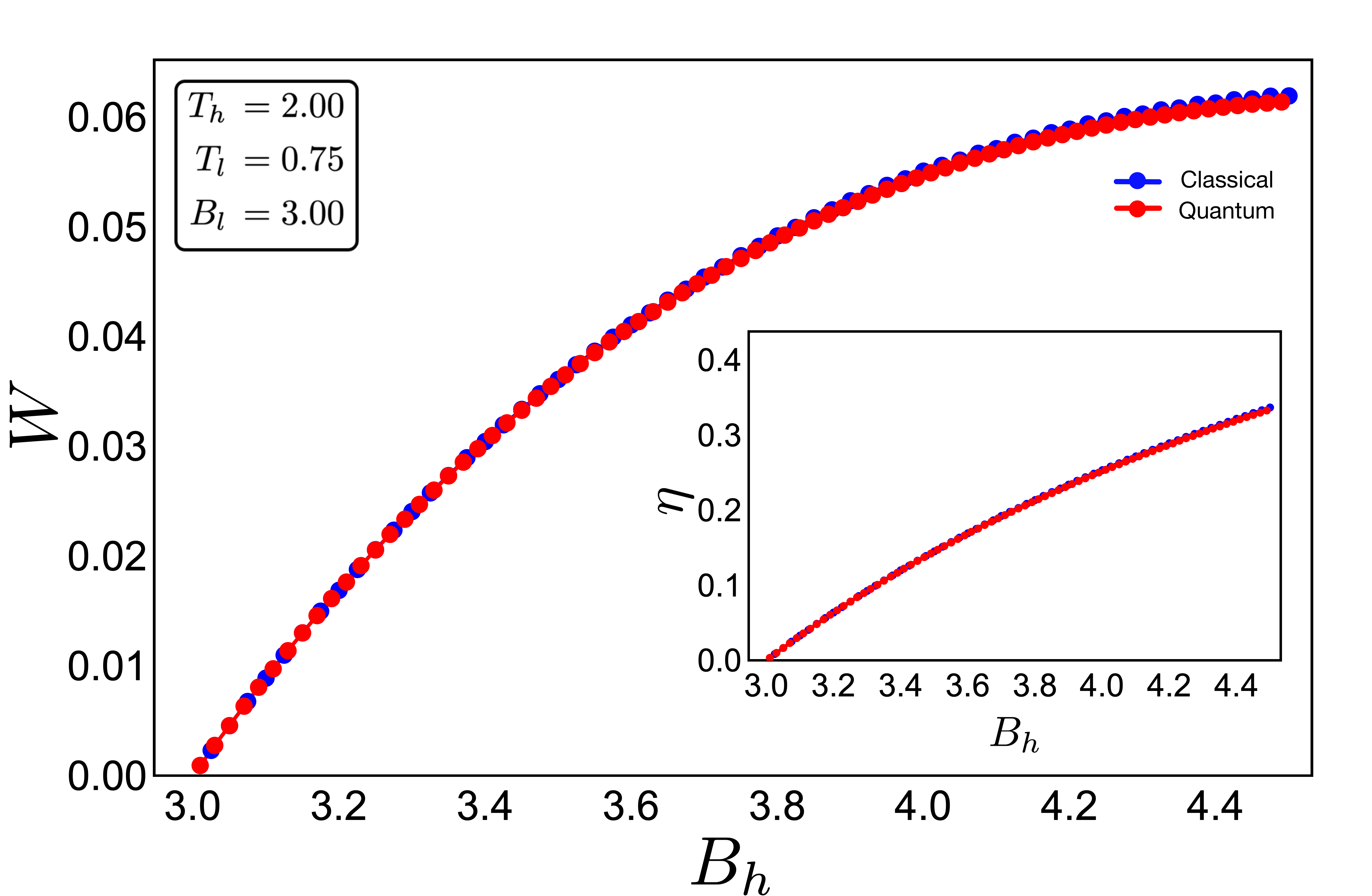}
     \caption{Work (in arbitrary units) and efficiency (inset) for the case of zone 1 (green rectangle in fig. \ref{quasistatictrajectories}) for a initial value of external magnetic field $B_{l}= 3.00$ and up to $B_{h}=4.50$. The hot an cold reservoirs are $T_{h}=2.00$ and $T_{l}=0.75$ respectively. The blue dotted line represents the classical performance, and the red dotted line represents the quantum one. }
     \label{work3levels}
   \end{figure}

To discuss the performance of the classic and quantum Otto cycle for the three-level system, first we obtain the diagram of temperature versus external field in fig. \ref{quasistatictrajectories}, when we apply the classical conditions described by equation (\ref{equality}). We select two zones where we will apply the cycle. The green  rectangle (zone 1) corresponds to a zone where the magnetic field and temperature  are related to each other in a linear way,  and the blue rectangle (zone 2) a region where the temperature and the external field have a non-linear relation between them. The work and efficiency for zone 1 is presented in fig. (\ref{work3levels}) for $T_{h}=2 $, $T_{l}=0.75$ with an starting value of the external field given by $B_{l}=3.00$ and the high magnetic field ($B_{h}$) is moving up to the value 4.5. As expected, in the three levels case, when the relation between the variables involved in the cycle is linear (under a classical adiabatic condition) the performance of the classical and quantum cycle it is the same.

   \begin{figure}[!ht]
       \includegraphics[width=0.48\textwidth]{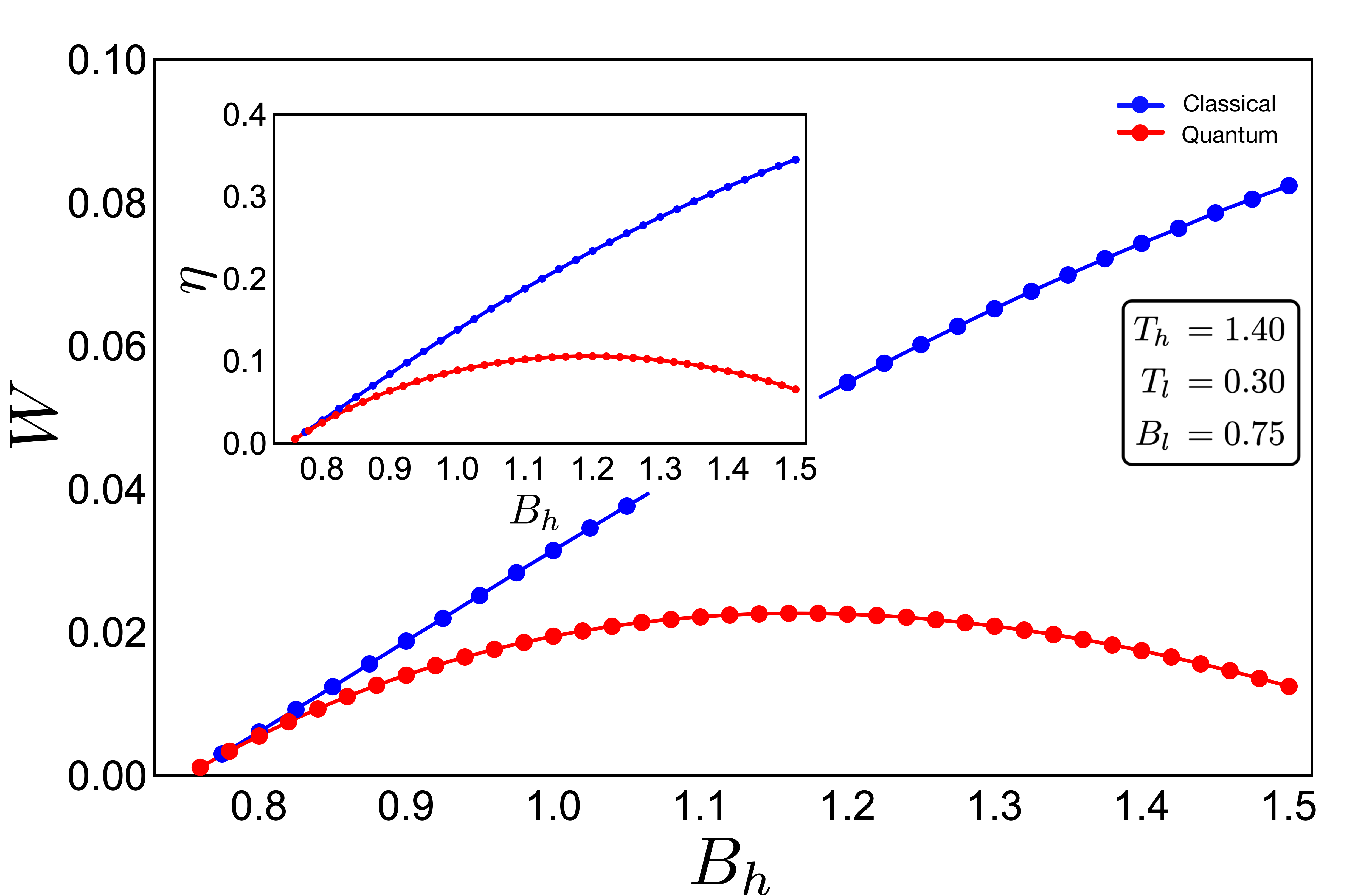}
     \caption{ Work (in arbitrary units) and efficiency (inset) for the case of zone 2 (blue rectangle in figure \ref{quasistatictrajectories}) for a initial value of external magnetic field $B_{l}= 0.75$ and up to $B_{h}=1.50$. The hot an cold reservoirs are $T_{h}=1.40$ and $T_{l}=0.30$ respectively. The blue dotted line represents the classical performance, and the red dotted line represents the quantum one.}
     \label{work3levels2}
   \end{figure}

If we analyse the work and efficiency for zone 2  in fig. (\ref{work3levels2}) where the parameters are given by  $T_{l}=0.30$ and $T_{h}=1.40$ and the external field is in the range of 0.75 up to 1.50, we observe a drastic decrease in work and efficiency for points very close to the initial field value $B_{l}$. Two things happen in that area which are fundamental to obtain these results. First, when analysing fig. \ref{quasistatictrajectories} in that zone,  a non-linear behavior between both variables is observed (as previously discussed), and second, observing the energy graph in fig. \ref{entropy3levels}(a), the three thermal populations of the levels involved are essential for the calculation of the total work extraction in the cycle. It is, therefore, that these results confirm that the inclusion of more energy levels to the systems causes the work and efficiency of the cycle to be lower than for a two-level scenario.

\section{Conclusions}
In this paper, we have studied and compared Otto's classical and quantum cycle showing that Otto's classical cycle extracts an equal or greater amount of work than its quantum counterpart. This is because, in the classical case, the system goes through four states of thermal equilibrium while for the quantum case it goes through only two. This is a general result of thermodynamics, valid for any working substance since, in a process of constant entropy, the minimum energy is that of thermodynamic equilibrium. Furthermore, this result is consistent with the maximum working postulate of thermodynamics since the Otto quantum cycle is an irreversible process, and the classical Otto cycle is reversible. 
We also study two particular cases in which the efficiency and the work in the classical and quantum approach are the same. These correspond to any two-level system, and also to the case in which a linear relationship is obtained between the temperature and the external field in the cycle during the adiabatic stages. This is because, in those two cases, the adiabatic trajectories conserve the thermal population of each quantum state during the process, thus satisfying the adiabatic condition required by the formulation of the Otto cycle in its quantum version.

\acknowledgments

F. J. P. acknowledges financial support from Conicyt (Chile) grant PAI77180015. P. V. and O.N. acknowledge  support from Conicyt (Chile) PIA/Basal AFB180001. O. N. acknowledge support from USM-DGIIP for Ph.D. fellowship. N.C.  acknowledges  support  from Conicyt,  Fondecyt  grant  no.  3200658.

\end{document}